\documentclass[prb, twocolumn, showpacs, amsmath, amssymb]{revtex4} 
\usepackage{multirow}
\usepackage{bbm}
\usepackage{MnSymbol}
\usepackage{bbold}
\usepackage{booktabs}
\usepackage{epsfig}
\usepackage{graphicx}
\usepackage{color}

\newcommand{\ix}[1]{\ensuremath{\text{#1}}} 
\newcommand{\K}{\ix{K}} 
\newcommand{\Ret}{\ix{ret}} 
\newcommand{\Adv}{\ix{adv}} 
\newcommand{\res}{\ix{res}} 
\newcommand{\hc}{\ensuremath{\text{H.c.}}} 
\newcommand{\abs}[1]{\ensuremath{\left| #1 \right|}} 

\DeclareMathOperator{\Tr}{Tr} 
\DeclareMathOperator{\Real}{Re} 
\DeclareMathOperator{\Imag}{Im} 
\renewcommand{\Re}{\Real}
\renewcommand{\Im}{\Imag}

\begin{document}

\title{
The interacting resonant level model in nonequilibrium: finite temperature effects}

\author{D.M.~Kennes} 
\author{V.~Meden} 

\affiliation{Institut f\"ur Theorie der Statistischen Physik, RWTH Aachen
  University and JARA---Fundamentals of Future Information
  Technology, 52056 Aachen, Germany}

\date{\today}

\begin{abstract}
We study the steady-state properties as well as the relaxation dynamics of the nonequilibrium interacting resonant level model at finite temperatures. It constitutes the prototype model of a correlated charge fluctuating quantum dot. The two reservoirs are held at different chemical potentials---the difference being the bias voltage---and different temperatures; we discuss the transport through as well as the occupancy of the single level dot. First, we show analytically that in the steady state the reservoir temperatures in competition with the other energy scales act as infrared cutoffs. This is rather intuitive but, depending on the parameter regime under consideration, leads to a surprisingly rich variety of power laws in the current as a function of the temperatures and the bias voltage with different interaction dependent exponents. Next we clarify how finite reservoir temperatures affect the dynamics. They allow to tune the interplay of the two frequencies characterizing the oscillatory part of the time evolution of the model at zero temperature. For the exponentially decaying part we disentangle the contributions of the level-lead hybridization and the temperatures to the decay rates. We identify a coherent-to-incoherent transition in the long time dynamics as the temperature is raised. It occurs at an interaction dependent critical temperature. Finally, taking different temperatures in the reservoirs we discuss the relaxation dynamics of a temperature gradient driven current.  
\end{abstract}


\maketitle


\section{Introduction}
\label{sec:introduction}
In recent years interacting nano-structures have attracted a great deal of both experimental as well as theoretical interest. They exhibit a variety of interesting many-body effects. Being equally fascinating and challenging to access theoretically, these systems are subject of a variety of studies (for a recent review see Ref. \onlinecite{Andergassen10}). An experimentally well controlled setup is given by a quantum dot with only a few electronic degrees of freedom contacted to a certain number of reservoirs. Due to the Coulomb repulsion these systems usually feature strong local electron correlations. A prominent consequence of those is the Kondo effect: if a nearly odd number of electrons reside on the quantum dot, spin fluctuations lead to a many-body resonance.\cite{Hewson93} Besides of spin fluctuating quantum dot setups a typical model of interest is the interacting resonant level model (IRLM), which is dominated by charge fluctuations.\cite{Schlottmann80,Filyov} The IRLM is given by a single level which is tunnel coupled to reservoirs (leads) of spinless fermions. A particle occupying the dot level interacts with the reservoir fermions located close to the dot by Coulomb repulsion. The level is coupled to an external gate, which allows to tune its energy. Here we study the IRLM in a minimal transport setup with two reservoirs.

Currently the research focus shifts from the equilibrium to the nonequilibrium physics of quantum dots. One can aim at the bias voltage driven steady-state (which is usually assumed to exist, see Refs.\ \onlinecite{Myohanen} and \onlinecite{Moldoveanu}) properties,\cite{flowequations,Rosch,rtrg,Herbert,Anders1,Anders2,PeterS,Fabian,Reinhold,Werner,Perfetto} or to be even more ambitious at the entire relaxation dynamics from transient to asymptotic.\cite{rtrg,Anders1,Anders2,Kehrein,Dirk,Perfetto} In the second case it is often assumed that at some time $t_0$ the quantum dot is coupled to the leads and thus relaxation from its initially prepared to its steady state sets in. In contrast to the well studied equilibrium properties, the nonequilibrium physics remains largely uncharted. In particular this holds if standard perturbative approaches fail to provide reliable results due to the presence of logarithmically divergent terms as it is the case in the IRLM.\cite{Schlottmann80,Karrasch10a} For such problems renormalization group (RG) approaches\cite{Herbert,nrg,Metzner11}  might succeed to describe the system under consideration. 

Recently a functional RG (FRG) approach\cite{Metzner11} was developed to tackle the nonequilibrium properties of interacting quantum dots\cite{Jacobs10} and tested successfully in its application to the IRLM.\cite{Karrasch10a,Karrasch10b,Kennes12} Combined with studies using alternative methods\cite{Doyon,Borda07,Boulat2,Mehta,Andergassen,Bran10a,Bran10b} this has led to a comprehensive understanding of the bias voltage driven steady state as well as the relaxation dynamics at vanishing temperature of the reservoirs. By far less is known about the physics at elevated temperatures.\cite{Doyon} Here we aim at closing this gap and present a detailed study of the nonequilibrium IRLM with finite reservoir temperatures using FRG. We discuss the steady state behavior as well as the dynamics at arbitrary (possibly asymmetric) temperatures in the reservoirs and for a general bias voltage. Our results are approximate, but controlled to leading order in the local electron-electron interaction; the RG procedure ensures that they are far superior to perturbation theory. Logarithms in the hopping amplitudes are resummed leading to power laws with interaction dependent exponents.\cite{Schlottmann80,Doyon,Karrasch10a,Karrasch10b,Kennes12,Borda08,Borda07,Boulat2,Filyov,Kashcheyevs} We uncover several interesting, partly surprising, temperature effects. First, we show analytically that in the steady state the temperatures serve as infrared cutoff scales to the renormalization of the tunnel couplings and as such compete with the other low-energy scales [see Eq.~(\ref{eq:renorm_ana})]. This is a rather intuitive finding, but depending on the parameter regime studied, the competition leads to a surprisingly rich variety of power laws in the dependence of the current on the temperatures and the voltage [see Eqs.~(\ref{noch}) to (\ref{auchnoch})] which can certainly not be guessed based on simple power-counting arguments or poor man's RG approaches. The interaction dependent exponents partly differ from the ones found at vanishing temperature. Our comprehensive study of the finite temperature effects thus reveals a very involved physics of the steady state current not reported on before. In the relaxation dynamics of the current and the occupancy varying the two temperatures allows to individually tune the amplitudes of the characteristic oscillatory terms and to vary the decay rates (see Fig.~\ref{fig:delT}). Crucially this cannot be achieved by varying the two tunnel couplings to the left and right leads; we disentangle the contributions of the temperatures and tunnel couplings to the decay rates. As an essential intermediate step of our analysis we present analytical expressions for the relaxation dynamics of the IRLM for vanishing interaction at arbitrary temperatures and bias voltage [see Eqs.~(\ref{eq:n_free(t)}) and (\ref{eq:J_free(t)})]. To the best of our knowledge those were not given before. We characterize a coherent-to-incoherent transition as the temperature is increased. At this the long time dynamics changes from being exponential with overlayed oscillations to being purely exponential. This transition takes place at a critical temperature which depends in a nontrivial way on the interaction strength. Finally, we investigate the time evolution of a current driven by a temperature gradient instead of a bias voltage, which for the IRLM constitutes a so far unexplored setup.

The rest of our paper is structured as follows. In Sect. \ref{sec:modelandmethod} we outline the model under consideration and additionally describe the FRG approach followed to tackle the problem at hand. Our result, first for the steady state analysis and than for the dynamics, are presented in Sect. \ref{sec:results}. We conclude our paper by a short summary in the final Sect. \ref{sec:summary}.

\section{Model and method}
\label{sec:modelandmethod}
 \subsection{Model} 
Our model is given by the Hamiltonian 
\begin{equation}
  H = H^\ix{dot} + \sum_{\alpha=L,R} [H^\res_\alpha + H^\ix{coup}_\alpha] \, .
\label{fullH}
\end{equation}
The dot part $ H^\ix{dot}$ consists of a linear geometry of three lattice sites with 
\begin{align}
  H^\ix{dot}_0 &= \epsilon n_2 - U\left(\frac{n_1}{2} + n_2 +
    \frac{n_3}{2}\right)   \nonumber
  \\
  & \hspace{3em} + (\tau_{12}d_1^\dagger d_2 +\tau_{23} d_2^\dagger d_3 + \hc) \, ,
\label{eq:singlepartinter}
  \\
  H^\ix{int} &= U (n_1 n_2 + n_2 n_3),\label{eq:twopartinter}
\end{align}
in standard second quantized notation.
Here $n_j = d_j^\dagger d_j$ is the occupancy operator of the spinless 
fermionic level $j$. The levels (sites) are connected locally through a hopping 
amplitude $\tau_{ij} >0$ and a 
Coulomb repulsion $U \geq 0$. The central site can be subject to a gate voltage allowing to tune the onsite energy $\epsilon$. The second term in the single-particle part 
of the Hamiltonian is added such that $ \epsilon=0 $ corresponds 
to half dot filling of the central dot site in equilibrium. 
The two leads $\alpha=L,R$ are modeled as noninteracting,
\begin{equation}
  H^\res_\alpha = \sum_{k_\alpha} \epsilon_{k_\alpha}
  c^\dagger_{k_\alpha} c_{k_\alpha} \; .  
\label{leadpartH}
\end{equation}
The left (right) lead is tunnel-coupled to side 1 (3) by
\begin{equation}
  H^\ix{coup}_\alpha =  \gamma_{\alpha} \sum_{k_\alpha}
  c^\dagger_{k_\alpha} d_{j_\alpha} + \hc \, ,
\end{equation}
with $j_L=1$ and $j_R=3$. For brevity we have not included time dependent parameters in $ H^\ix{dot}$ and $H^\ix{int}$ which can also be treated by our approach as outlined in Ref.~\onlinecite{Kennes12b} for reservoir temperatures $T_{L/R}=0$. 

We assume that the system is prepared in a product density matrix state $\rho$ at time $t_0=0$, which describes a situation naturally arising when the coupling between the dot and the reservoirs vanishes for $t<0$. Furthermore, the reservoirs are in grand canonical equilibrium with $T_\alpha=1/\beta_\alpha$ and chemical potentials centered around zero $\mu_L=-\mu_R=V/2\geq 0$, that is
\begin{align}
  \rho(t=0) &= \rho_0 =  \rho^\ix{dot}_0 \otimes
  \rho^\res_{L,0}  \otimes \rho^\res_{R,0},
  \\
  \rho^\res_{\alpha,0} &= e^{-(H^\res_\alpha - \mu_\alpha
    N_\alpha)/T_\alpha} / \Tr e^{-(H^\res_\alpha - \mu_\alpha N_\alpha)/T_\alpha},
\end{align}
where $N_\alpha = \sum_{k_\alpha} c^\dagger_{k_\alpha} c_{k_\alpha}$. We choose units with the Boltzmann constant $k_\ix{B} = 1$, $\hbar = 1$, and one electron charge $e=1$. Finally, we assume that the statistical operator $\rho^\ix{dot}_0$ at $t=0$ describes an initially empty quantum dot. This mostly studied relaxation protocol is not the one realized in experiments. In those the dot is initially coupled and in equilibrium with unbiased (in voltage and temperature) leads. At $t=0$ a bias (in voltage and/or temperature) is turned on.\cite{Stefanucci04} Effects of the initial correlations present in such a setup can be studied with our FRG approach. This was done in Ref.~\onlinecite{Kennes12b} at vanishing temperature. It turned out that the initial correlations die out exponentially with the typical decay rate. Needless to say they do not effect the steady state. We are confident that the finite temperature effects in the relaxation dynamics discussed here equally appear in the presence of initial correlations.   

We aim at a limit in which the model is equivalent to the field theoretical IRLM.\cite{Schlottmann80,Doyon,Borda08,Borda07,Boulat2,Mehta,Filyov,Kashcheyevs} This is called the scaling limit and can be achieved by coupling the first and third site much stronger to their respective reservoirs than to the central site. It is then that the first and third dot sites can effectively be incorporated into the leads\cite{Karrasch10a,Kennes12} resulting in a single level (central site) tunnel coupled to and interacting with the reservoirs. We are interested in the universal physics of the IRLM which is independent of the details of the leads band structure. Therefore we study structureless reservoirs (wide band limit) with hybridization
\begin{align}
  {\Gamma_\alpha} &= \pi D_\alpha |\gamma_{\alpha}|^2
\end{align}
much larger than all other energy scales and constant density of states 
\begin{equation}
  D_\alpha(\epsilon) = D_\alpha e^{-\delta \abs{\epsilon}},
\end{equation}
with $\delta \rightarrow 0^+$ assuring convergence of the energy integrals. For simplicity of depiction we will use symmetric couplings to the reservoirs of site one and three $\Gamma_\alpha=\Gamma$. In summary the limit $\tau_{ij}, |\epsilon|, |U|, V,|T_\alpha| \ll \Gamma$ establishes that the model is equivalent to the field theoretical IRLM which is the prototype model of a quantum dot dominated by correlated charge fluctuations.

 \subsection{Method}  To analyse the behavior of the IRLM at finite reservoir temperatures we extend the FRG methods described in Ref.~\onlinecite{Karrasch10a} for the nonequilibrium steady state and Ref.~\onlinecite{Kennes12} for the time evolution. We here give an outline of our approach and discuss the essential new steps. For a more detailed description we refer the reader to the aforementioned publications. 
 
To tackle the nonequilibrium problem at hand we employ the Keldysh formalism.\cite{HaugJauho,Rammer} In a two step procedure we first include the influence of the reservoirs on the noninteracting dot and secondly consider the effect of the two-particle interaction both in form of self-energy contributions.
The corresponding Dyson equations are given as Eqs.~(12), (13), (25), (29), (30), and (31) in Ref.~\onlinecite{Kennes12}.
The reservoir self-energy $\Sigma_\res$ is calculated exactly, while the contribution of the two-particle interaction is determined via the approximate FRG approach. It was successfully applied to equilibrium\cite{Metzner11} and nonequilibrium\cite{Jacobs10,Karrasch10a,Kennes12} transport through correlated quantum dots before. In FRG one introduces a flow parameter $\Lambda$ in the noninteracting part of the propagation. We specify the cutoff procedure by coupling an auxiliary structureless reservoir to each of our three dot sites via hybridization $\Lambda$. Taking the derivative of the generating functional with respect to the $\Lambda$ yields an exact infinite hierarchy of flow equations for the vertex functions which, as the only approximation of our method, is truncated to a given order. The cutoff-free problem is recovered after integrating from $\Lambda=\infty$ (where the vertices are known analytically) down to $\Lambda=0$. Here we use the lowest truncation order; the resulting flow equation for the interaction part of the self-energy is given in Eqs.~(44) and (45) of Ref.~\onlinecite{Kennes12}. The flowing self-energy matrix elements can be interpreted as flowing single-particle parameters. Those are the onsite energy of site one and three $\epsilon^{\prime,\Lambda}$ (renormalize equally in our truncation), the onsite energy of the central site $\epsilon^{\Lambda}$ as well as the hopping between sites one and two $ \tau^{\Lambda}_{12}$ and between two and three $ \tau^{\Lambda}_{23}$ (generically renormalize differently in nonequilibrium). The problem to be solved at the end of the RG flow corresponds to an effective noninteracting one with 
renormailzed single-particle parameters. 

The described truncation already allows to obtain a comprehensive understanding of the nonequilibrium physics of the IRLM at zero temperature and for small to intermediate interactions: the logarithmically divergent terms present in lowest-order perturbation theory are resummed consistently leading to RG-renormalized hopping amplitudes featuring generic power laws with interaction-dependent exponents, which are correct to leading order in the interaction.\cite{Karrasch10a} For the time evolution FRG leads to terms exponentially decaying in time with interaction dependent decay rates as well as power-law corrections $t^{-\kappa}$ with $U$-dependent exponent\cite{Kennes12} $\kappa$ also found in an alternative RG procedure.\cite{Karrasch10b,Andergassen} 
 
\noindent \emph{Steady State}---We employ the cutoff via auxiliary reservoirs featuring infinite temperature (instead of zero temperature as in the $T_\alpha=0$-study of Ref.~\onlinecite{Karrasch10a}). We explicitly showed that infinite and zero temperature in the auxiliary reservoirs give the same results\cite{Kennes12,DanteMaster} (the former implies technical simplifications) exemplifying the robustness of the cutoff procedure. The reservoir Keldysh self-energy for the present cutoff procedure takes the form (in lattice site space)
\begin{equation}
\Sigma^\K_\res(\omega)=4 i \Gamma \begin{pmatrix}
f_L(\omega)-1/2&0&0\\
0&0&0\\
0&0&f_R(\omega)-1/2 
\end{pmatrix}
\end{equation}
with $f_\alpha$ being the Fermi function of reservoir $\alpha$. The Keldysh Green function is then determined via
\begin{equation}
G^{\K,\Lambda}(\omega)=G^{\Ret,\Lambda}(\omega)\Sigma^\K_\res(\omega)G^{\Adv,\Lambda}(\omega)
\end{equation}
since the Keldysh self-energy from the two-particle interaction vanishes in our truncation. We can use unchanged expressions for retarded and advanced Green functions as compared to the $T=0$ case;\cite{Karrasch10a} the temperature does not enter in these quantities explicitly. The single-scale propagator $S^{\K,\Lambda}(\omega)$ is related to the Keldysh Green function by Eq.~(45) of Ref.~ \onlinecite{Kennes12}. As the latter depends on the $T_\alpha$ we have to account for finite temperature changes in the flow equations. Those read
\begin{align}
\partial_\Lambda \tau_{12}^\Lambda&=\frac{iU}{4\pi}\int S^{\K,\Lambda}_{12}(\omega)d\omega,\;\;\;\; \tau_{12}^{\Lambda\to\infty}=\tau_{12}\label{eq:flow_tau12}\\
\partial_\Lambda \tau_{23}^\Lambda &=\frac{iU}{4\pi}\int S^{\K,\Lambda}_{23}(\omega)d\omega,\;\;\;\; \tau_{23}^{\Lambda\to\infty}=\tau_{23}\\
\partial_\Lambda \epsilon^\Lambda &=-\frac{iU}{4\pi}\int \left[S^{\K,\Lambda}_{11}+S^{\K,\Lambda}_{33}\right](\omega)d\omega,\;\;\;\; \epsilon^{\Lambda\to\infty}=\epsilon\\
\partial_\Lambda \epsilon^{\prime\Lambda} &=-\frac{iU}{4\pi}\int S^{\K,\Lambda}_{22}(\omega)d\omega,\;\;\;\; \epsilon^{\prime\Lambda\to\infty}=0\label{eq:flow_eps'}.
\end{align}

\noindent \emph{Time evolution}---For the time evolution we generalize the method outlined in Ref.~\onlinecite{Kennes12} to finite temperatures. Again we aim at the changes induced by the reservoir Keldysh self-energy, which enter the Keldysh Green function $G^K(t,t)$ and the single scale propagator $S^K(t,t)$. For this it is necessary to evaluate integrals of the type
\begin{equation}
\int dt \frac{e^{-a t}}{\sinh(\pi T_\alpha t)}
\end{equation}
which can be done analytically in form of polygamma $\Psi(n,x)$ and hypergeometric functions ${}_2\mathcal{F}_1(a,b,c,z)$.\cite{abramowitz} Since the formulas are rather lengthy, they are given in the Appendix.

\section{Results}
\label{sec:results}
\subsection{Steady State} 
\noindent \emph{Temperature as a cutoff scale}---At this point one can proceed with a numerical solution of the flow equations \eqref{eq:flow_tau12}-\eqref{eq:flow_eps'}. We postpone this and first report on analytical results. To obtain those we suppress the renormalization of the central onsite energy $\epsilon^{\Lambda}$ which is $\mathcal{O}(U^2)$ (remind that in our truncation we only control terms to order $U$) as well as $\epsilon^{\prime\Lambda}$, which always appears in combination with the much larger scale $\Gamma$. This gives
\begin{align}
\partial_\Lambda \tau_{12}^\Lambda=&-\frac{U\Gamma}{\pi}\int\limits_{-\infty}^\infty\partial_\Lambda^*\left[G^{\Ret,\Lambda}_{11}(\omega)(f_L(\omega)-1/2)G^{\Adv,\Lambda}_{12}(\omega)\right]d\omega\notag\\
&-\frac{U\Gamma}{\pi}\int\limits_{-\infty}^\infty\partial_\Lambda^*\left[G^{\Ret,\Lambda}_{13}(\omega)(f_R(\omega)-1/2)G^{\Adv,\Lambda}_{32}(\omega)\right]d\omega\\
\partial_\Lambda \tau_{23}^\Lambda=&-\frac{U\Gamma}{\pi}\int\limits_{-\infty}^\infty\partial_\Lambda^*\left[G^{\Ret,\Lambda}_{21}(\omega)(f_L(\omega)-1/2)G^{\Adv,\Lambda}_{13}(\omega)\right]d\omega\notag\\
&-\frac{U\Gamma}{\pi}\int\limits_{-\infty}^\infty\partial_\Lambda^*\left[G^{\Ret,\Lambda}_{23}(\omega)(f_R(\omega)-1/2)G^{\Adv,\Lambda}_{33}(\omega)\right]d\omega
\end{align}
as the remaining flow equations. We here introduced the star differential operator $\partial^*_\Lambda$ which acts only on the free 
Green function $ \hat G^{0,\Lambda} $, not on $ \Sigma^\Lambda $, in the series expansion $ \hat G^{\Lambda}= \hat G^{0,\Lambda}+\hat G^{0,\Lambda}\Sigma^\Lambda \hat G^{0,\Lambda}+\dots$.
From these expressions we can extract the corresponding flow equations of the level-lead hybridization $\Theta_{ij}=|\tau_{ij}|^2/\Gamma$ suppressing all terms $\mathcal{O}(1/\Gamma^{2})$
\begin{align}
\partial_\Lambda \Theta_{12}^\Lambda=&\frac{U}{2\pi\Gamma}\partial_\Lambda^*\int\limits_{-\infty}^\infty \bigg[\tanh\left(\frac{\beta_L(\omega-\mu_L)}{2}\right)\notag\\&\times\frac{\Theta_{12}^\Lambda}{\omega-\epsilon-i(\Lambda+\Theta_{12}^\Lambda+\Theta_{23}^\Lambda)}+\mbox{c.c.}\bigg] d\omega\label{eq:Ren_Theta12}\\
\partial_\Lambda \Theta_{23}^\Lambda=& \partial_\Lambda \Theta_{12}^\Lambda(1\leftrightarrow 3, L\leftrightarrow R).
\end{align}
In a good approximation one can neglect the renormalization of $\Theta_{12}$ and $\Theta_{23}$ in the denominator---denoted as approximation 1 in the following---and integrate Eq. \eqref{eq:Ren_Theta12} to
\begin{align}
&\partial_\Lambda \Theta_{12}^\Lambda=-\frac{U}{\pi\Gamma}\Theta_{12}^\Lambda\notag\\&\times\Re\bigg\{\frac{\beta_L}{\pi}\Psi\left(1,\frac{1}{2}-\frac{i}{\pi}\frac{\beta_L}{2}\left[\epsilon-\mu_L+i(\Lambda+\Theta_{12}+\Theta_{23})\right]\right)\bigg\},\label{eq:flow_approx1}
\end{align} 
with $\Psi(1,x)$ being the trigamma function. Additionally, setting $\Psi(1,1/2+x)\approx 1/(2/\pi^2+x)$ for all $\Re[x]>0 $ (ensuring the correct value at $x=0$ as well as asymptotic behavior) for the right hand site of the flow equation---denoted as approximation 2 in the following---finally yields
\begin{equation}
\partial_\Lambda \Theta_{12}^\Lambda\approx -\frac{U}{\pi\Gamma}\Theta_{12}^\Lambda \frac{2(2 T_L/\pi+\Lambda+\Theta_{12}+\Theta_{23})}{(2T_L/\pi+\Lambda+\Theta_{12}+\Theta_{23})^2+[(\epsilon-\mu_L)/2]^2}.\label{eq:cutoffbeforeint}
\end{equation}
After reintroducing the ultraviolet cutoff\cite{comment1} $\Gamma$ one can integrate this equation analytically and obtains 
\begin{align}
\frac{\Theta_{12}^{\Lambda=0}}{\tau_{12}^2}\sim \begin{cases}
\begin{matrix}
\left(\tau_{12}^2\right)^{-2U/(\pi\Gamma)+\mathcal{O}(U^2)}\;\;\;\; |\epsilon-\mu_L|,T_{L}\ll T_K\ll\Gamma \\
V^{-2U/(\pi\Gamma)+\mathcal{O}(U^2)}\;\;\;\; |\epsilon|,T_K,T_{L}\ll V\ll\Gamma \\|\epsilon|^{-2U/(\pi\Gamma)+\mathcal{O}(U^2)}\;\;\;\; V,T_K,T_{L}\ll |\epsilon|\ll\Gamma \\
T_L^{-2U/(\pi\Gamma)+\mathcal{O}(U^2)}\;\;\;\; |\epsilon-\mu_L|,T_K\ll T_{L} \ll\Gamma \label{eq:renorm_ana}
 \end{matrix}
\end{cases}
\end{align}
with $\tau=\tau_{12}+\tau_{23}$ and $T_K$ being the universal equilibrium low energy scale of the model\cite{Schlottmann80,Doyon,Borda08,Borda07,Boulat2,Mehta,Filyov,Kashcheyevs} defined via the charge susceptibility\cite{Karrasch10a}
\begin{equation}
 \chi=\left.\frac{d\bar n}{d\epsilon}\right|_{\epsilon=0}=-\frac{2}{\pi T_K}.
\end{equation} 
Equations analogous to \eqref{eq:flow_approx1}-\eqref{eq:renorm_ana} are obtained for the renormalization of $\Theta_{23}^{\Lambda=0}$ by replacing $(1\rightarrow 3,L\rightarrow R)$. It is apparent [see Eq. \eqref{eq:renorm_ana}] that the reservoir temperatures provide infrared cutoffs. Equation \eqref{eq:cutoffbeforeint} indicates the interplay of the different cutoff scales $\epsilon-\mu_\alpha,\Theta_{12}+\Theta_{23},T_\alpha$ entering in the denominator explicitly. We find that the left (right) hybridization is cut off by the temperature of the left (right) reservoir only. This invokes important consequences for the behavior of the current outlined in more detail in the next section. Different cutoff scales for the left and right hybridization have been observed before (at $T_\alpha=0$); it is $|\epsilon-\mu_L|$ ($|\epsilon-\mu_R|$) and not $V=\mu_L-\mu_R$ which enters as a cutoff for $\Theta_{12}$ ($\Theta_{23}$).\cite{Karrasch10b} 

\begin{figure}[t]
  \centering \includegraphics[width=\linewidth,clip]{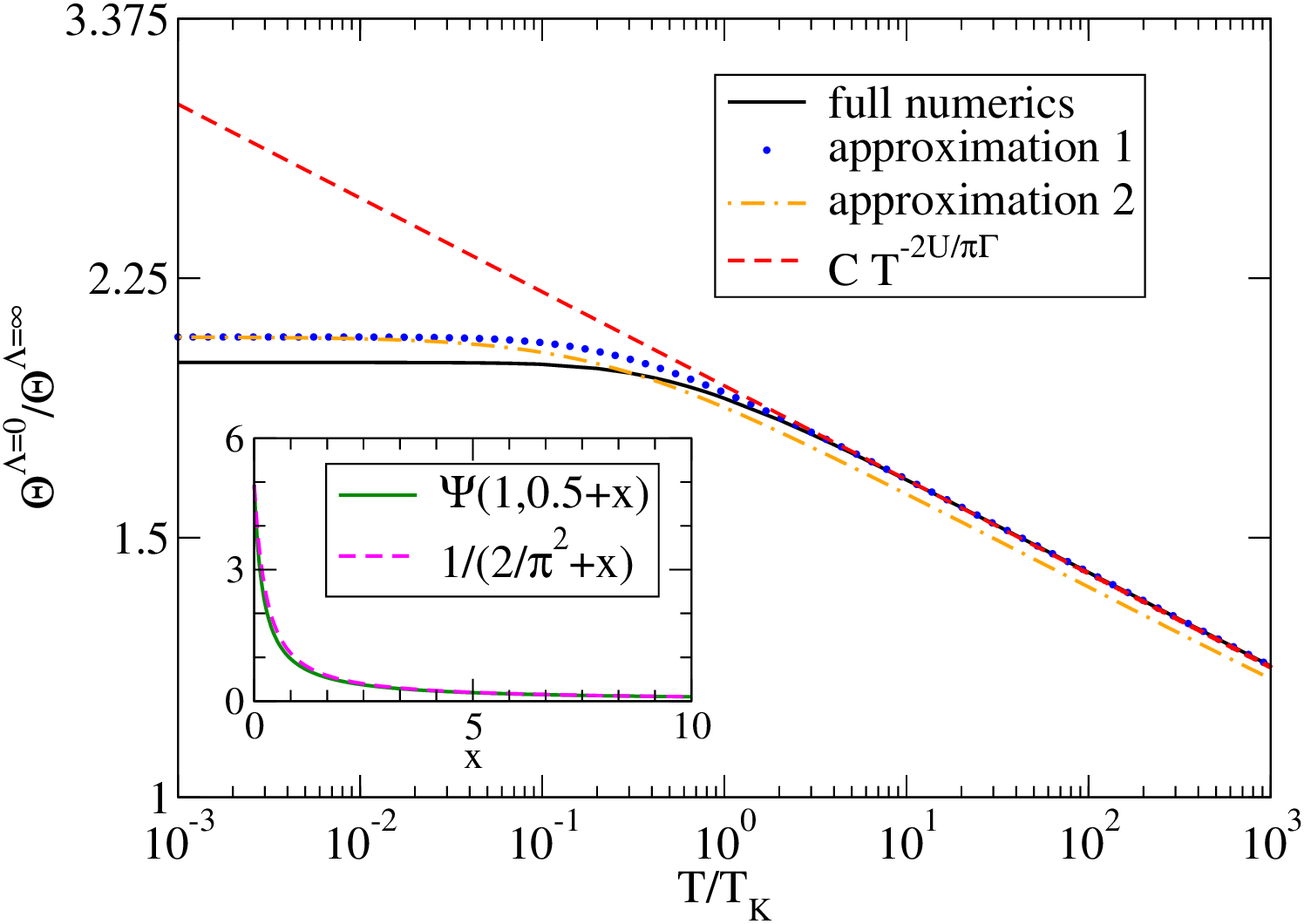}
  \caption{(Color online) Renormalized line width $\Theta^{\Lambda=0}$ as a function of temperature $T$. Parameters are chosen as $\epsilon/T_K=V/T_K=0$, $\tau_{12}/\Gamma=\tau_{23}/\Gamma=0.0025$, $T=T_L=T_R$ and $U/\Gamma=0.1$. Since left and right lead are chosen symmetrically $\Theta_{12}=\Theta_{23}=\Theta$ holds. Notice the double logarithmic scale, which indicates the power law behavior for large $T$. The inset shows the replacement of the trigamma function used in approximation 2.}
  \label{fig:Anacut}
\end{figure}

We next elaborate on the quality of the approximations $1$ and $2$. To this end we take the full numerical solution of the flow equations (the only approximation being the truncation) and step by step apply approximations 1 and 2. A comparison for the renormalized line width $\Theta^{\Lambda=0}$ as a function of temperature is shown in Fig.~\ref{fig:Anacut}. For the entire temperature range ($T_L = T_R = T$ and $\tau_{12} = \tau_{23} = \tau$ for simplicity) the results of approximations 1 and 2 agree well with the numerical solution. In particular for $T \gtrapprox T_K$, the power law Eq. \eqref{eq:renorm_ana} with the correct exponent is reproduced on each level of approximation.

\noindent \emph{Observables}---Within our truncation the effect of the interaction can be cast into effective single-particle parameters. To gain analytical insights it is therefore sufficient to derive expressions for the occupancy and the current at vanishing interaction. One can simply supplement the noninteracting formulas with the effective parameters obtained at the end of the flow to incorporate the interaction. We focus on the current which is the most interesting observable in our transport setup. We start from the Meir-Wingreen form\cite{meir,comment2}
\begin{align}
J_L=&4\Gamma^2\int d\omega[f_L(\omega)-f_R(\omega)]|G^{\Ret,\Lambda=0}_{13}|^2\notag\\
\approx& 4\Theta^{\Lambda=0}_{12}\Theta^{\Lambda=0}_{23}\int d\omega[f_L(\omega)-f_R(\omega)]\notag\\&\phantom{4\Theta^{\Lambda=0}_{12}\Theta^{\Lambda=0}_{23}\int d\omega}\times\left|\frac{1}{i\omega-i\epsilon-\Theta^{\Lambda=0}_{12}-\Theta^{\Lambda=0}_{23}}\right|^2,
\end{align}  
where terms $\mathcal{O}(U^2)$ (which we do not controll in any case) as well as $\mathcal{O}(1/\Gamma^2)$ (scaling limit) were neglected in the second step. The integral can be solved and yields
\begin{align}
J_L=&4\frac{\Theta^{\Lambda=0}_{12}\Theta^{\Lambda=0}_{23}}{\Theta^{\Lambda=0}_{12}+\Theta^{\Lambda=0}_{23}}\sum\limits_{\alpha=L,R}  {s(\alpha)}\notag\\&\times\Im\left[\Psi\left(\frac{1}{2}-\frac{\beta_\alpha}{2\pi}(i(\mu_\alpha-\epsilon)-(\Theta^{\Lambda=0}_{12}+\Theta^{\Lambda=0}_{23})) \right)\right]\label{eq:I_approx1}
\end{align}
with $\Psi(x)$ being the digamma function, $s(L)=-1$ and $s(R)=1$. Combining this expression with the flow Eq.  \eqref{eq:flow_approx1} and its analogue for $\Theta_{23}^{\Lambda}$ allows us to obtain a comprehensive analytical understanding of the steady-state current.  

In the ohmic regime (in which the current depends linearly on the voltage), for $T_L = T_R =T$, $\tau_{12} = \tau_{23} = \tau$, and at low $T$ where $T_K\gg T,V,|\epsilon|$ the $T$-dependence of the current was earlier studied in  Ref.~\onlinecite{Doyon} using a field theoretical approach. In this limit we find 
\begin{equation}
\Theta_{12}^{\Lambda=0}=\Theta_{23}^{\Lambda=0}=\Theta_{12}\left(\frac{\Gamma}{T_K}\right)^{\frac{2U}{\Gamma \pi}}\exp\left[-\frac{2U}{\Gamma \pi}\left(\frac{\bar T^2}{6}-\frac{7\bar T^4}{60}\right)\right],
\end{equation}
where we neglected terms $\mathcal{O}(\bar T^6)$ with $\bar T=\pi T/T_K$. Plugging the renormalized hybridizations into the current formula \eqref{eq:I_approx1} yields
\begin{equation}
\frac{J_L}{V}=1-\frac{1}{3}\bar T^2+\left(\frac{7}{15}-\frac{2U}{9\Gamma\pi}\right)\bar T^4
\end{equation}
up to $\mathcal{O}(\bar T^6)$ and $\mathcal{O}(U^2)$. The ratio $R=g_4/g_2^2$ of the prefactors of $\bar T^2$ ($g_2$) and $\bar T^4$ ($g_4$) is given by $R=21/5-2U/\Gamma\pi$, which is the same result as found in Ref.~\onlinecite{Doyon} (see also Ref.~\onlinecite{Boulat}).

At $T_{L/R}=0$, for left-right symmetric tunnel couplings, $\epsilon=0$ and for sufficiently large bias voltages Eq. \eqref{eq:renorm_ana} leads to a power-law suppression  $J_L\sim V^{-\nu}$ of the current with $\nu=-\frac{2U}{\pi\Gamma}$.\cite{Doyon,Borda07,Boulat2} It was later shown\cite{Karrasch10b,Andergassen} that relaxing the second and/or third requirement leads to a more complicated form of the current-voltage characteristics. Allowing for a left-right asymmetry in the reservoir's temperatures introduces another level of complexity. We break it up by considering the symmetric case $T_\alpha=T\gg |\epsilon-\mu_\alpha|,T_K,$ first. In this regime the flow of both level-lead couplings is cut off by the temperature and the current reads
\begin{align}
J_L\sim T^{-\frac{2U}{\pi\Gamma}-1},
\label{noch}
\end{align}
as long as $V\neq 0$ or $ \epsilon \neq 0$. We emphasize that the exponent is different from the above given $\nu$ appearing in the the voltage dependence. To the best of our knowlegde it has not been found before. For $V=0$ and $ \epsilon = 0$, $J_L=0$ independent of the temperature difference in the reservoirs. This follows right away from half filling. In the regime $T_L\gg T_R\gg  |\epsilon-\mu_\alpha|,T_K,$  
\begin{align}
J_L\sim (T_L T_R)^{-\frac{2U}{\pi\Gamma}-1}\frac{c_2 T_L+T_R}{T_L^{-\frac{2U}{\pi\Gamma}}+c_1T_R^{-\frac{2U}{\pi\Gamma}}},
\end{align}
where $c_1$ stems from the asymmetry in the renormalization of the left and right hoppings (the exact form of $c_1$ is irrelevant for the following discussion) and $c_2=-(\epsilon-V/2)/(\epsilon+V/2)$. Off resonance $(\epsilon \neq V/2)$ the current is given by
\begin{align}
J_L\sim (T_L T_R)^{-\frac{2U}{\pi\Gamma}-1}\frac{T_L}{T_L^{-\frac{2U}{\pi\Gamma}}+c_1T_R^{-\frac{2U}{\pi\Gamma}}},
\end{align}
but on resonance\cite{Karrasch10b,Andergassen} $(\epsilon = V/2)$ it changes to
\begin{align}
J_L\sim (T_L T_R)^{-\frac{2U}{\pi\Gamma}-1}\frac{T_R}{T_L^{-\frac{2U}{\pi\Gamma}}+c_1T_R^{-\frac{2U}{\pi\Gamma}}}.
\end{align}
Similarly one finds for $T_L\gg V \gg  T_R,|\epsilon|,T_K,$ 
\begin{align}
J_L\sim \frac{(T_L V)^{-\frac{2U}{\pi\Gamma}}}{T_L^{-\frac{2U}{\pi\Gamma}}+c_1V^{-\frac{2U}{\pi\Gamma}}}.
\label{auchnoch}
\end{align}
This exemplifies how the temperature significantly changes the qualitative behavior of the  current, although it enters the renormalization of the hybridizations in an (``simple'') intuitive way (as an infrared cutoff). We emphasize that the subtle interplay of the lead temperatures and the voltage revealed by the above equations can only be uncovered using a method (such as ours) which allows for an unbiased RG-like treatment of multiple energy scales.  

\subsection{Time evolution} 
\noindent \emph{The noninteracting case}---To gain a comprehensive understanding of the effect of finite reservoir temperatures on the relaxation dynamics of the IRLM it is advantageous to first study the noninteracting case. We derive closed analytical expressions for the time dependence of the current and occupancy, which to the best of our knowledge were not presented before. Based on those the effect of $T_\alpha >0$ on the relaxation rates and characteristic oscillations can be worked out in detail. A comparison with the $U>0$ results obtained numerically by solving the FRG flow equations and computing $\bar n(t)$ and $J_L(t)$  then allows to assess the correlation effects.  The correlation physics is particular prominent and transparent for $|\epsilon\pm V/2|\gg T_K$. It was earlier shown that for $T_\alpha=0$ in this regime the relaxation dynamics for $U>0$ and sufficiently large times  can be obtained by replacing the time-independent bare single-particle parameters in analytical $U=0$ expressions by the time-averaged renormalized ones.\cite{Kennes12} We show that the same holds for finite reservoir temperatures. This constitutes a second reason for the 
$U=0$ expressions of $\bar n(t)$ and $J_L(t)$ providing the basis of our understanding also of the interacting model.

We introduce the hybridization of the single-level model $\tilde\Gamma_{ij}$. To connect to the three site model in the scaling limit one has to set
$\tilde \Gamma_{ij}={|\tau_{ij}|^2}/{\Gamma}.$ For simplicity we focus on the case of symmetric couplings to the reservoirs $\tilde \Gamma_{ij}=\tilde \Gamma$. The occupancy $\bar n(t)$ with the initial value $\bar n_0$ reads (for expressions of the occupancy in terms of Green functions see Ref.~\onlinecite{Kennes12})
\begin{widetext}
\begin{equation}
\begin{split}
\bar n(t)=&\frac{1}{2}-\frac{1}{2}e^{-4\tilde \Gamma t}(1-2\bar n_0)+\frac{1}{2}\sum\limits_\alpha T_\alpha \Im\bigg\{\frac{1}{\pi T_\alpha}\Psi\left(\frac{-(i(\epsilon-\mu_\alpha)-2\tilde\Gamma-\pi T_\alpha)}{2\pi T_\alpha}\right)-e^{-4\tilde \Gamma t}\frac{1}{\pi T_\alpha}\Psi\left(\frac{-(i(\epsilon-\mu_\alpha)+2\tilde\Gamma-\pi T_\alpha)}{2\pi T_\alpha}\right)\\
&-\frac{2e^{(i(\epsilon-\mu_\alpha)-2\tilde\Gamma-\pi T_\alpha)t}}{(i(\epsilon-\mu_\alpha)-2\tilde\Gamma-\pi T_\alpha)}{}_2\mathcal{F}_1\left[1,\frac{-(i(\epsilon-\mu_\alpha)-2\tilde\Gamma-\pi T_\alpha)}{2\pi T_\alpha},\frac{-(i(\epsilon-\mu_\alpha)-2\tilde\Gamma-\pi T_\alpha)}{2\pi T_\alpha}+1,e^{-2\pi T_\alpha t} \right]\\
&+\frac{2e^{(i(\epsilon-\mu_\alpha)-2\tilde\Gamma-\pi T_\alpha)t}}{(i(\epsilon-\mu_\alpha)+2\tilde\Gamma-\pi T_\alpha)}{}_2\mathcal{F}_1\left[1,\frac{-(i(\epsilon-\mu_\alpha)+2\tilde\Gamma-\pi T_\alpha)}{2\pi T_\alpha},\frac{-(i(\epsilon-\mu_\alpha)+2\tilde\Gamma-\pi T_\alpha)}{2\pi T_\alpha}+1,e^{-2\pi T_\alpha t} \right]\bigg\} . \label{eq:n_free(t)}
\end{split}
\end{equation}
\end{widetext}
Interestingly, the rate $4\tilde \Gamma$ appearing in the first exponential term remains independent of the temperatures. For the time dependence of the current we find (for equations expressing the current in terms of Green functions we again refer to Ref.~\onlinecite{Kennes12})
\begin{widetext}
\begin{equation}
\begin{split}
\frac{J_L(t)}{\tilde \Gamma}=&1-2\bar n(t)+4T_L\frac{
    \Im\left[\Psi\left(\frac{i(\mu_L - \epsilon) + 2\tilde\Gamma + \pi T_L}{2 \pi T_L}\right)\right]}{2 \pi T_L} \\
    &-4T_L\Re\bigg\{i \frac{e^{-(i(\mu_L - \epsilon) + 2\tilde\Gamma + \pi T_L)t}}{i(\mu_L - \epsilon) + 2\tilde\Gamma + \pi T_L}{}_2\mathcal{F}_1\left[1,\frac{i(\mu_L - \epsilon) + 2\tilde\Gamma + \pi T_L}{2\pi T_\alpha},\frac{-(i(\epsilon-\mu_\alpha)+2\tilde\Gamma-\pi T_\alpha)}{2\pi T_\alpha}+1,e^{-2\pi T_\alpha t}\right] \bigg\} .  \label{eq:J_free(t)}
\end{split}
\end{equation}
\end{widetext}
Up to the above mentioned first time dependent term in $\bar n(t)$, which remains unchanged if $T_\alpha$ is increased, the dynamics is affected by the reservoir temperatures [e.g. compare to Eqs. (109)-(111) of Ref. \onlinecite{Andergassen}]. The long-time behavior of the remaining terms in $\bar n(t)$ and $J_L(t)$ changes from being governed by an exponential relaxation in combination with a power-law correction for $T=0$\cite{Karrasch10b,Andergassen,Kennes12} to an infinite series of exponential terms with temperature dependent rates $2\tilde \Gamma +2n\pi T_\alpha$ and $n\in \mathbbm{N}$ at $T> 0$. The frequencies characterising the oscillatory part of the behavior of $\bar n(t)$ and $J_L(t)$ are the same as for $T=0$:\cite{Kennes12,Andergassen} in general one finds the frequencies $|\epsilon\pm V/2|$, where the amplitude of $|\epsilon+ V/2|$ ($|\epsilon-V/2|$) is suppressed in the regime $|\epsilon\pm V/2|\gg T_K$ in the current $J_L$ ($J_R$).

Obviously the reservoir temperatures do not play the same role as the hybridisation (remind that $\tilde \Gamma_{12} = \tilde \Gamma_{23} =\tilde \Gamma$). In contrast to the latter, which enters in every relaxation rate of Eqs.  \eqref{eq:n_free(t)} and \eqref{eq:J_free(t)}, the temperature allows to tune the influence of the different terms with respect to each other; the rates dependent solely on either $T_L$ or on $T_R$. Furthermore, even with asymmetric couplings to the left and right reservoir, that is generalizing Eqs. \eqref{eq:n_free(t)} and  \eqref{eq:J_free(t)}, only the sum of the hybridisations enters the decay rates. The temperatures instead allow to tune the relaxation rates imprinted by the left and right reservoir independently. 

\noindent \emph{The interacting case}---In the developed FRG approach we determine the Keldysh and retarded Green functions of the interacting system. Given those we can numerically compute the observables as outlined in Ref.~\onlinecite{Kennes12}. We postpone this and first consider the time dependent renormalization of the single-particle parameters. When combined with Eqs.~\eqref{eq:n_free(t)} and  \eqref{eq:J_free(t)} the latter give valuable insights in the effect of the two-particle interaction. We carefully checked that the numerical parameters underlying the (numerical) solution of  the flow equations (for details see Ref.~\onlinecite{Kennes12}) are chosen such that the results are completely converged on the scales of all figures shown. In fact, it is one of the advantages of our approach (particular, compared to purely numerical ones) that for small to intermediate $U$ higly accurate results can be obtained for arbitrarily large times.

\begin{figure}[t]
\centering  
   \includegraphics[width=0.9\linewidth,clip]{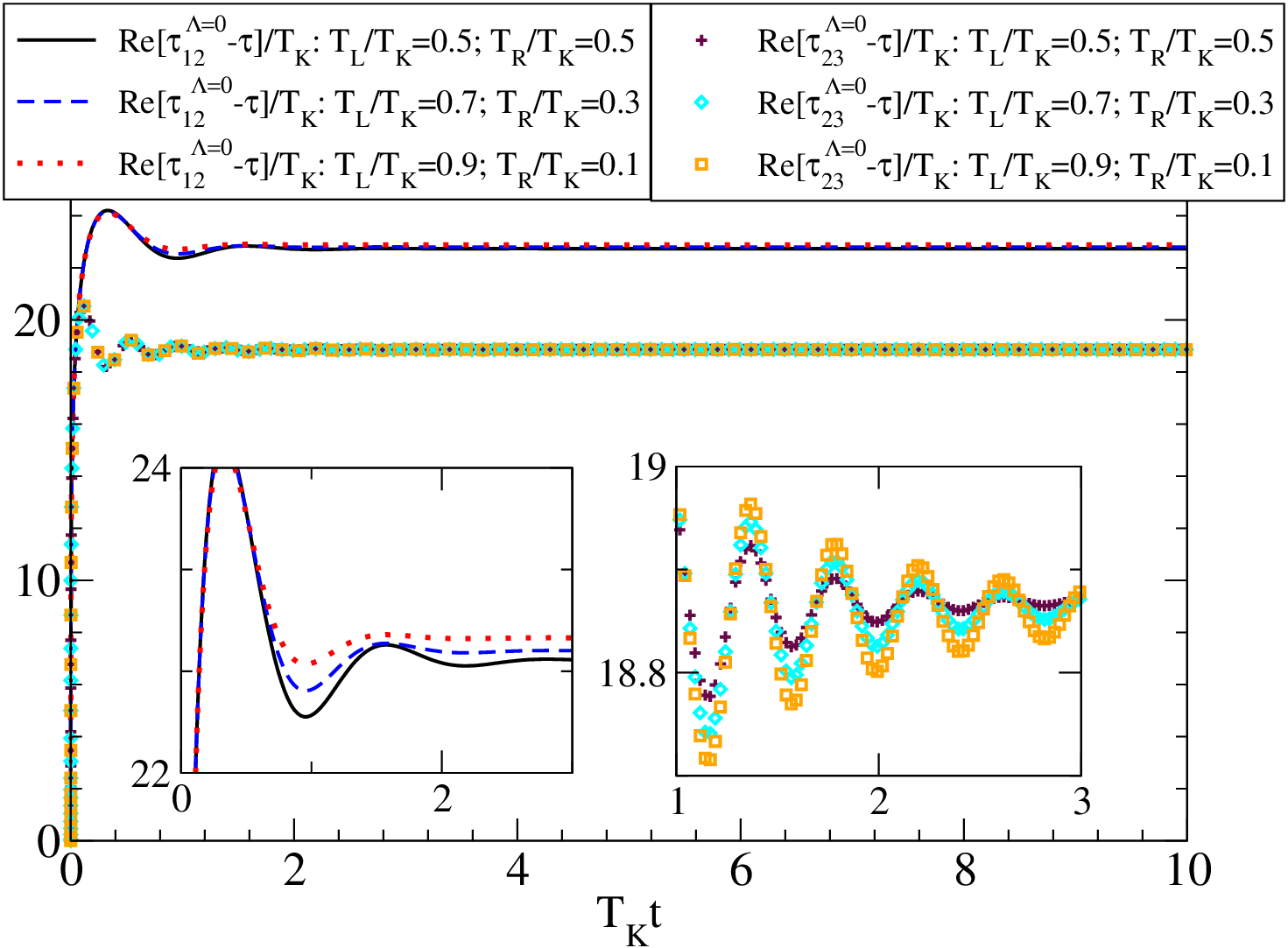}
   \caption{(Color online)  Time dependence of the renormalized hoppings $ \tau_{12}^{\Lambda=0} - \tau =  \Sigma^{\Ret,\Lambda=0}_{12} \in {\mathbbm C}$ between sites 1 and 2, 
and $ \tau_{23}^{\Lambda=0} - \tau = \Sigma^{\Ret,\Lambda=0}_{23} \in {\mathbbm C}$ between sites 2 and 3. The parameters are: $\tau_{12} = \tau_{23} = \tau$ with $\tau/\Gamma=0.0025$, $\epsilon/T_K=V/T_K=10$, and $ U/\Gamma=0.2$. The insets show a zoom into the regime where one can most clearly distinguish the temperature's effect on the oscillations' amplitude.}
   \label{fig:renormth}
\end{figure}

For arbitrary $T_\alpha$ and the relevant times $T_K t \gtrapprox \tau/\Gamma$ the renormalization of the onsite energy $ \epsilon^{\Lambda=0}(t)-\epsilon =  \Sigma^{\Ret,\Lambda=0}_{22}(t)$ is of order $U^2$ and can safely be neglected.
The time dependence of the renormalized hopping amplitudes is shown in Fig. \ref{fig:renormth} (for $\tau_{12} = \tau_{23} = \tau$). It depends only weakly on the temperature for the considered case of large $|\epsilon\pm V/2|$. As for $T=0$\cite{Kennes12} the effective hopping amplitudes quickly (on a  scale $T_K t \sim \tau/\Gamma$) start to oscillate around their steady state values with frequencies $|\epsilon-V/2|$ and $|\epsilon+V/2|$ for $\tau_{12}^{\Lambda=0}$ and $\tau_{23}^{\Lambda=0}$ respectively. Note that due to the left-right asymmetry induced by the voltage and different reservoir temperatures the renormalized level-lead couplings generically differ even for the bare ones being equal. Increasing the temperature of the left (right) reservoir suppresses oscillations in the renormalized $\tau_{12}^{\Lambda=0}$ ($\tau_{23}^{\Lambda=0}$); this will manifest as an interesting tunability in 
observables (see Fig.~\ref{fig:delT}).  For the sake of simplicity we only show the real part at this point, but the same holds for the  much smaller imaginary one as well. 

The analysis of the time dependence of the renormalized single-particle parameters in the limit $|\epsilon\pm V/2|\gg T_K$ shows that to interpret the results obtained by the numerical solution of the FRG flow equations and numerical computation of the observables for $T_K t \gtrapprox \tau/\Gamma$ one can refer to Eqs. \eqref{eq:n_free(t)} and  \eqref{eq:J_free(t)} with the bare parameters replaced by the time-averaged renormalized ones. More precisely we have to extend these expressions to the case of left-right asymmetric hybridzations. For briefness we do not give those here but note that the ratio of the hybridizations enters as prefactors in the different terms. 

 \begin{figure}[t]
\centering
\includegraphics[width=0.9\linewidth,clip]{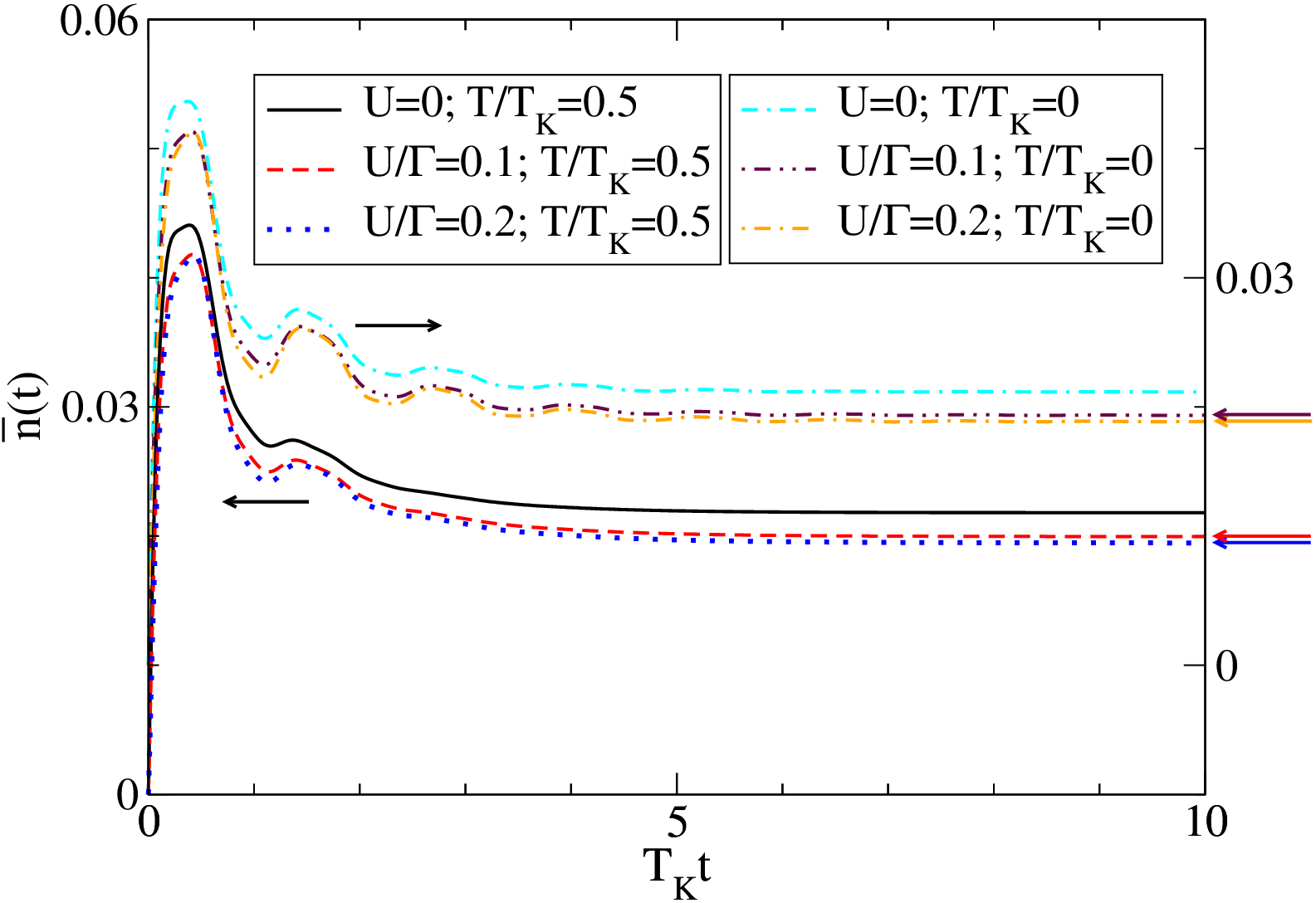}  
 \caption{(Color online) FRG data for the time evolution of the central-site occupancy. The parameters are the same as in Fig.\ \ref{fig:renormth} and symmetric temperatures $T_\alpha=T$. The arrows right to the graph indicate the steady-state values obtained by the nonequilibrium steady-state functional RG discussed earlier.} 
   \label{fig:Timeevolutionocc}
\vspace{.5cm}
\centering
 \includegraphics[width=0.9\linewidth,clip]{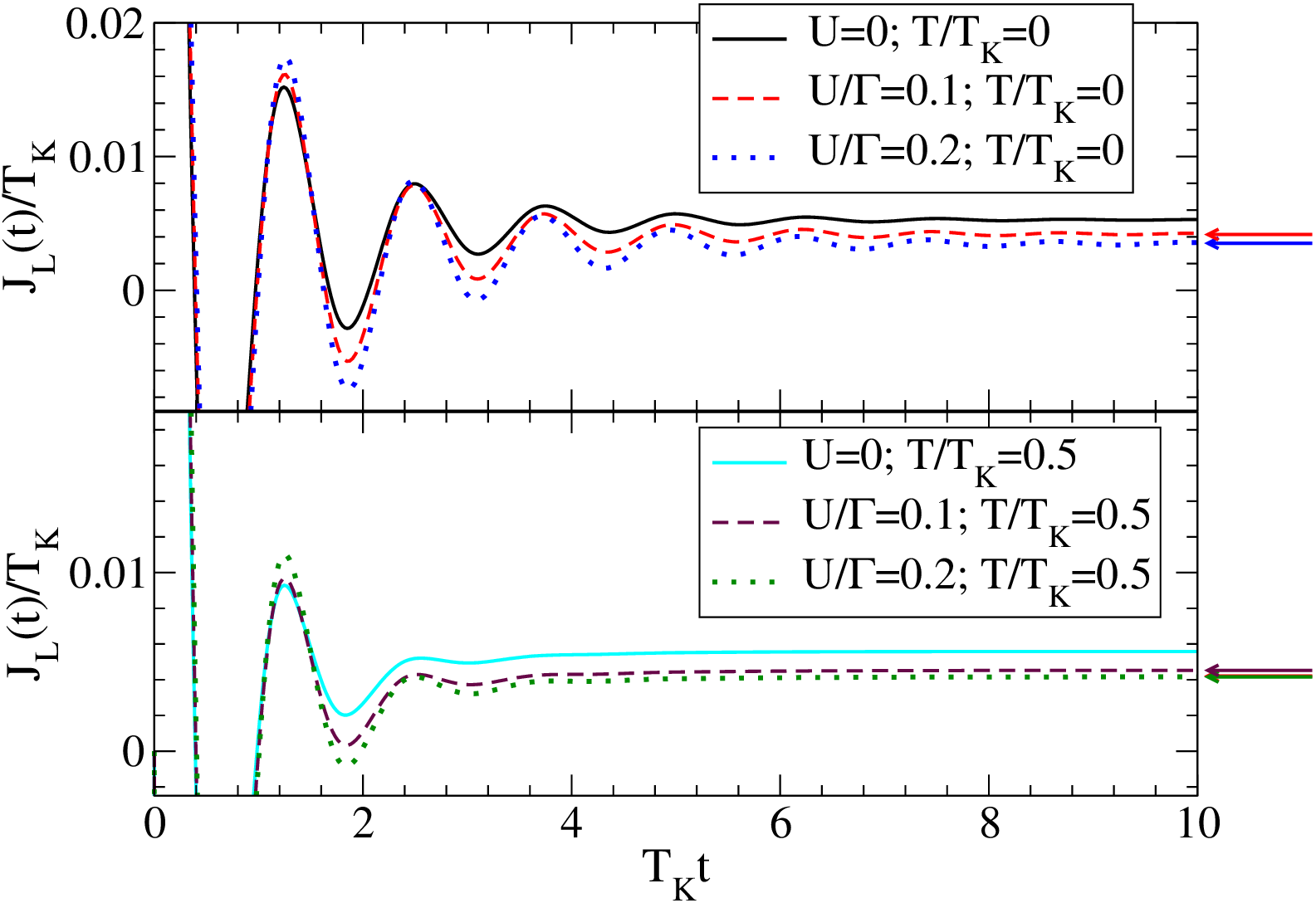}
\caption{(Color online) The same as in Fig.\ \ref{fig:Timeevolutionocc} but for the current leaving the left reservoir.} 
   \label{fig:Timeevolutioncur}
\end{figure} 

We now discuss our numerical FRG results for $\bar n(t)$ and $J_L(t)$ at $U>0$. To understand those we use the just established relation to the noninteracting dynamics. Figures \ref{fig:Timeevolutionocc} and \ref{fig:Timeevolutioncur} show the time evolution of the occupancy and the current in the limit $|\epsilon\pm V/2|\gg T_K$ for different $U$. Increasing the temperature drastically suppresses the amplitude of the oscillatory terms. The quality factor is decreased by the increased decay rate. One can tune the quality factor of the two frequencies $|\epsilon\pm V/2|$ independently of each other, which is not possible via the hybridization. Figure~\ref{fig:delT} illustrates this point for the time dependence of the occupancy. For vanishing temperature gradient $\Delta T=0$ it shows a superposition of the two frequencies $|\epsilon\pm V/2|$ (bottom curve in Fig.~\ref{fig:delT}). Raising the temperature gradient pronounces the contribution of the frequency $|\epsilon+V/2|$ belonging to the colder reservoir until the signal appears almost sinusoidal (overlayed by an exponential decay) at maximum temperature gradient (top curve in Fig.~\ref{fig:delT}). The inset shows the imaginary part of the numerical Laplace transform (normalized to its largest value) of the occupany for the smallest and largest temperature gradient. The positions of the frequencies $|\epsilon\pm V/2|$ are indicated by the vertical arrows. With increasing $ \Delta T$ the feature at $|\epsilon-V/2|$ is suppressed, while the one at $|\epsilon+V/2|$ is enhanced.
 
\begin{figure}[t]
\centering  
   \includegraphics[width=0.85\linewidth,clip]{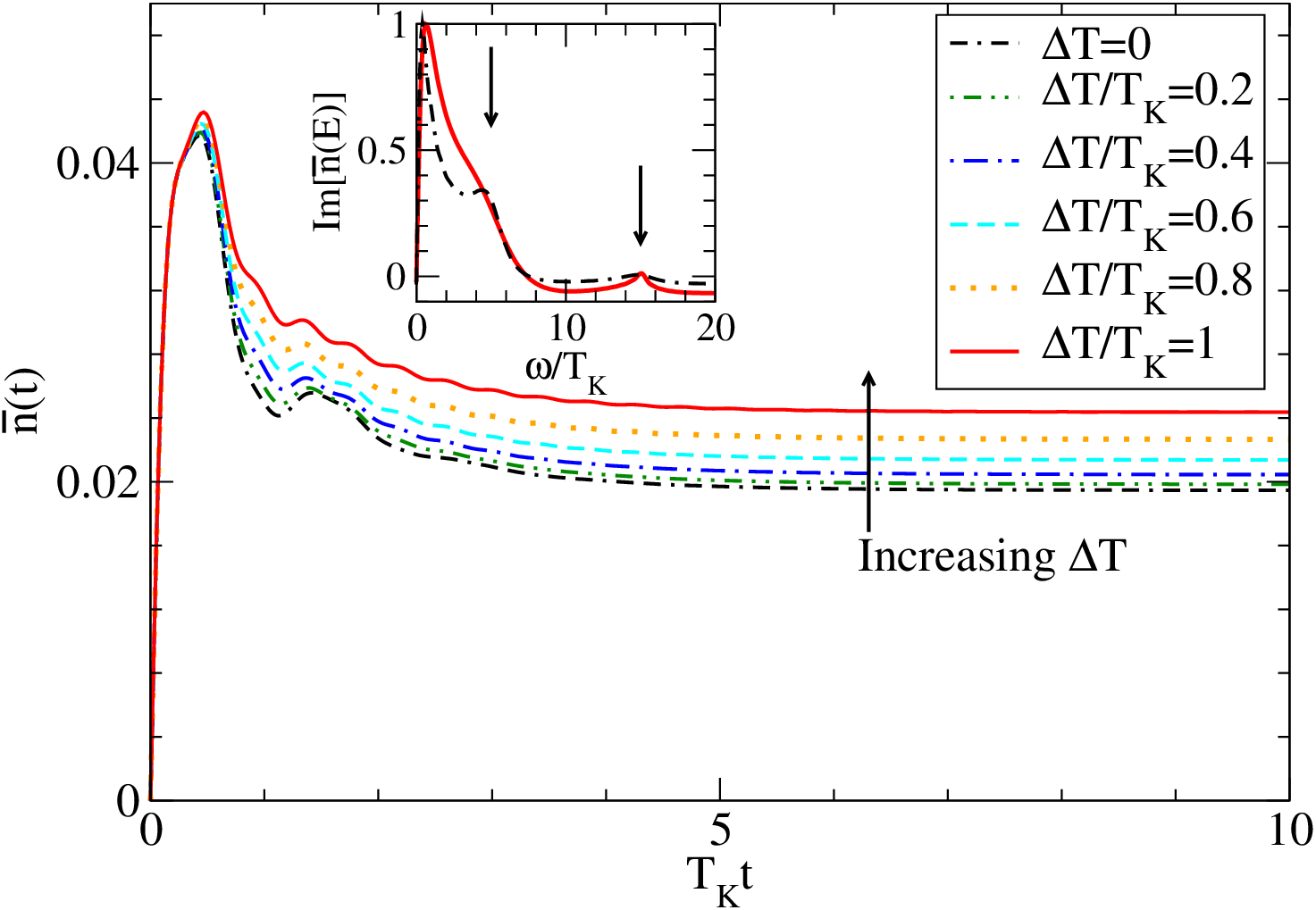}
   \caption{(Color online) The same as Fig.~\ref{fig:Timeevolutionocc} with $U/\Gamma=0.2$, but featuring a temperature gradient $T_L=T+\Delta T/2$ and $T_R=T-\Delta T/2$ for $T=0.5T_K$. We find how the temperature gradient pronounces the frequency $|\epsilon+V/2|$ belonging to the colder reservoir. This can also be found in the inset, which shows the numerical Laplace transform with the positions of the frequencies $|\epsilon\pm V/2|$ indicated by arrows. }
   \label{fig:delT}
\end{figure}

Next we study the far-from-equilibrium case $V\gg T_K,\epsilon=0$, which on general grounds is the most intriguing one. More precisely we consider the limit $V\gg T_K,T$ with $T_L=T_R=T$ and $\epsilon=0$. At $\epsilon=0$ the expressions for the time evolution are particularly simple and for sufficiently large times $\pi Tt\gg 1$ and small interactions we obtain  
  \begin{equation}
  J_L(t)=J_{\ix{stat}}+e^{-4 \Theta^{\Lambda=0} t}-2T\Im\left[\frac{4e^{(-iV/2-2\Theta^{\Lambda=0}-\pi T)t}}{i V+ 4\Theta^{\Lambda=0}+2\pi T}\right],
\label{eq:Phaseform}
\end{equation}  
with the renormalized steady-state values $\Theta^{\Lambda=0}=\Theta^{\Lambda=0}_{12}=\Theta^{\Lambda=0}_{23}$ and the stationary current $J_{\ix{stat}}$. We note that for initially equal $\Theta_{ij}$ and $\epsilon=0$ also the renormalized hybridizations remain equal even for $V \neq 0$. From Eq.  \eqref{eq:Phaseform} we can read off a coherent-to-incoherent transition in the long-time dynamics at temperature 
\begin{equation}
\pi T_c = 2\Theta^{\Lambda=0}.\label{eq:Phaset}
\end{equation}
The long time dynamics switches from being exponential with an overlayed oscillation (second term of Eq. \eqref{eq:Phaseform} being the dominate one at large times) for $T< T_c$ to purely (monotonic) exponential relaxation (first term of Eq. \eqref{eq:Phaseform} being the dominate one at large times) for $T>T_c$. To determine $T_c$ we can take the expression for $\Theta^{\Lambda=0}$ corresponding to the considered parameter regime $V\gg T_K,T$ of  Eq. \eqref{eq:renorm_ana}  $\Theta^{\Lambda=0} \sim \frac{\tau^2}{\Gamma}\left(\frac{\Gamma}{V}\right)^{2U/\pi/\Gamma}$. With this we obtain 
\begin{equation}
T_c=\frac{2\tau^2}{\pi\Gamma}\left(\frac{\Gamma}{V}\right)^{2U/\pi/\Gamma},
\end{equation}  
for the critical temperature at which the crossover from coherent to incoherent behavior occurs. Relaxing the condition of equal temperatures in the left and right reservoirs but keeping $V\gg T_K,T_L,T_R$ the transition is found at $\min (T_L,T_R)=T_c$ with the same $T_c$ as defined above. 

\begin{figure}[t]
\centering  
   \includegraphics[width=0.9\linewidth,clip]{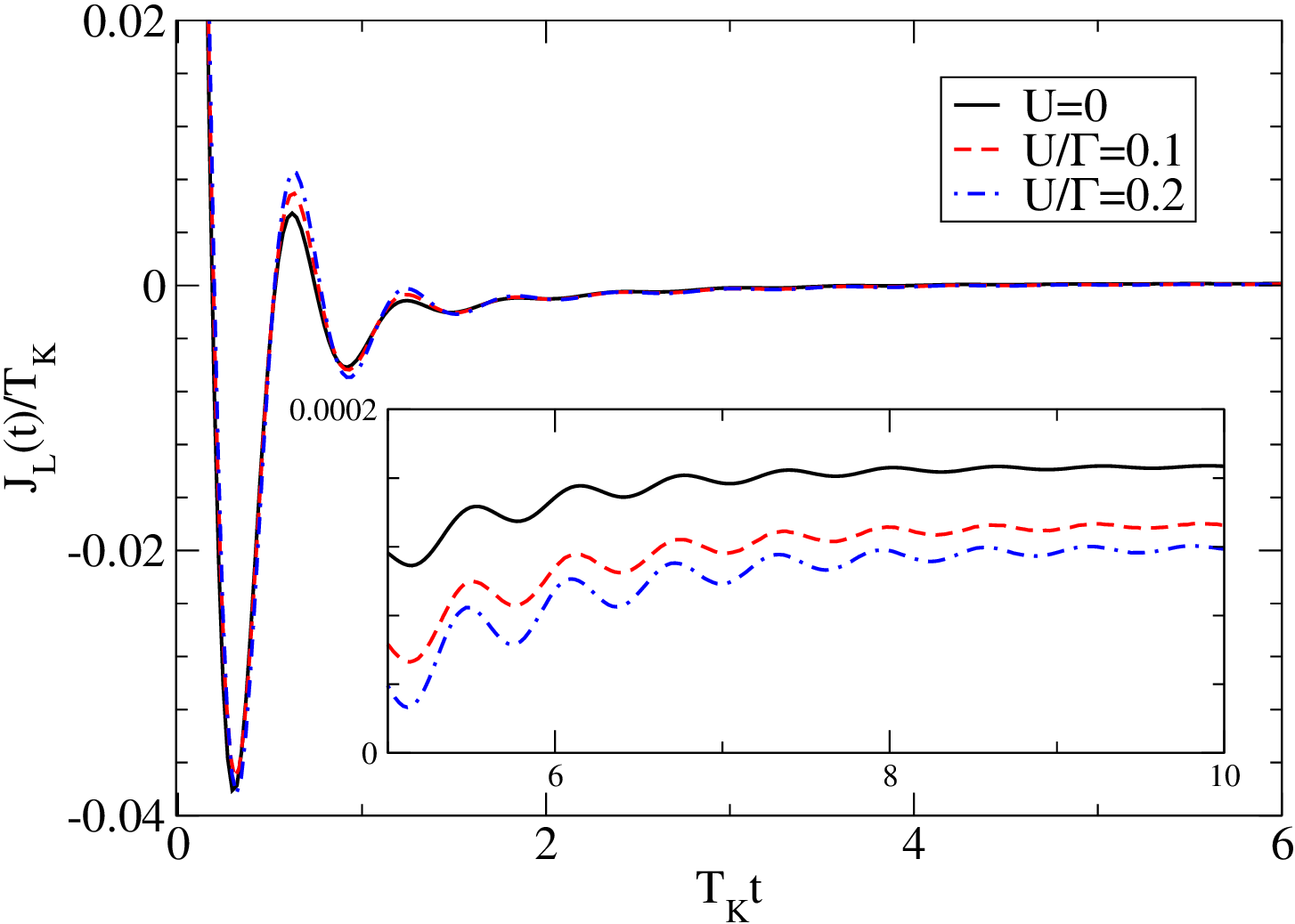}
   \caption{(Color online) Time evolution for the temperature gradient induced current leaving the left reservoir $J_L(t)$. The parameters read $\tau/\Gamma=0.0025$, $\epsilon/T_K=10$ and $V/T_K=0$. We choose $T_L=T_K$ and $T_R=0$ (choosing $0< T_R\ll T_L$ does not alter the qualitative behavior).}
   \label{fig:JdelT}
\end{figure}

We close the discussion by considering the time dependence of a current which is not induced by a finite bias voltage but by a finite temperature gradient accross the quantum dot. To obtain a nonvanishing steady-state value of the current one needs to choose $\epsilon\neq 0$. Figure \ref{fig:JdelT} shows the current for this parameter regime. Note that the only frequency in the time evolution is given by $\epsilon$ as $V=0$. The interaction strength enhances the amplitude of the oscillations, but decreases the steady-state current (if measured with respect to $T_K$). This investigation opens the road to thermal and thermoelectric transport studies which will be presented elsewhere.

\section{Summary}
\label{sec:summary}
We extended the functional RG analysis of the nonequilibrium interacting resonant level model in the scaling limit, which is the prototype of a simple charge fluctuating quantum dot, to the case of arbitrary reservoir temperatures. Our study includes all time regimes from transient to asymptotic with the ultimate limit of the steady state. All results developed are controlled to leading order in the interaction, but known to be far superior to simple perturbation theory as our RG procedure implies the resummation of logarithmically divergent terms to power laws.

First we discussed the role of temperature as an RG cutoff competing with the other energy scales of the model in the steady state. Although the temperature enters the renormalization of the hybridization in an intuitive way, the current was shown to exhibit a vast variety of different power laws (in temperature and bias voltage) with interaction dependent exponents for different parameter regimes. We than clarified the role of finite temperatures for the relaxation dynamics. We focused on the most transparent case $|\epsilon\pm V/2|\gg T_K,T_L,T_R$ in which the physics of the IRLM can be understood by interpreting the interacting system as an effective free one featuring renormalized parameters. The finite reservoir temperatures $T_\alpha$ (only) partly act similar to the hybridizations; increasing $T_\alpha$ increases the relaxation rate. Our detailed analysis shows that this obvious effect does not exhaust the role of temperature in the relaxation dynamics. We have shown explicitly that one can use the $T_\alpha$ to suppress the oscillatory contributions to the occupancy with the frequencies $|\epsilon\pm V/2|$ individually which is impossible by varying the hybridizations. We charcterized a coherent-to-incoherent transition in the long-time relaxation dynamics as the temperature is increased. The critical temperature at which the transition occurs depends in a nontrivial way on the two-particle interaction. Finally, we discussed a current, which is not driven by a bias voltage, but by a temperature gradient. We emphasize that, similar to the $T=0$ case, our approach can also be used to tackle explicitly time-dependent Hamiltonians, which constitutes a field of current interest. A direction of future research which opens up directly from the present work is the field of thermal and thermoelectric transport. In this context quantum dots are promising candidates for highly efficient energy conversion devices and thus have recently gained sizeable interest.\cite{Murphy08T,Kubala10T,Costi10T,Andergassen11} 


\section*{Acknowledgments}
We thank Sabine Andergassen and Dirk Schuricht for very valuable discussions as well as Frank Reininghaus for providing a numerical implementation of the complex trigamma function. This work was supported by the DFG via FOR 723. 
\appendix*
\setcounter{equation}{0}
\section*{Appendix}
\label{appen}

In the main part of this work it was discussed, that one needs to calculate the Keldysh Green function $G^K(t,t)$. It is determined by the integral
\begin{widetext}
\begin{equation}
\int \limits_{t_n}^{t_{n+1}}ds_1\int\limits_{t_m}^{t_{m+1}}ds_2 G^{\Ret}(t_{n+1},s_1)\Sigma^\K_\res(s_1,s_2) G^{\Adv}(s_2,t_{m+1}) 
\label{intappendix}
 \end{equation}
with $\Sigma^{\K}_\res(s_1,s_2)$ given by

\begin{equation}
\Sigma^\K_\alpha(t',t)= \Gamma \begin{pmatrix}
   - T_L e^{- i \mu_L(t'-t)}

  \sum_\pm \frac{1}{\sinh[\pi T_L(t'-t\pm i
    \delta)]},&0&0\\
0&0&0\\
0&0& - T_R e^{- i \mu_R(t'-t)}

  \sum_\pm \frac{1}{\sinh[\pi T_R(t'-t\pm i
    \delta)]} 
\end{pmatrix}.
\end{equation}
\end{widetext}
The retarded Green function can be evaluated as in the $T=0$ case via
\begin{equation}
G^{\Ret}_{ij}(t,t')=-i \sum\limits_{l=1}^3\text{Res}_{ij,l}e^{-i\omega_l(t-t')},
\end{equation}
with $ \text{Res}_{ij,l} $ and $ \omega_l $ being the residues and the poles of 
\begin{equation}
\label{eq:forresiduesandpoles}
\frac{1}{\omega-\left(\tilde h_0^{\rm dot} + \Sigma^\Ret_{\res}-i\Lambda+ \Sigma^\Ret_{\bar t}\right)}.
\end{equation}
Introducing the time dependent effective parameters
$\epsilon^{\prime\Lambda} = \Sigma^\Ret_{\bar t,11}-U/2 $,
$\epsilon^{\Lambda} = \epsilon+\Sigma^\Ret_{\bar t,22}-U$,
$\tau_{12}^\Lambda = \tau+\Sigma^\Ret_{\bar t,12}$ and 
$\tau_{23}^\Lambda = \tau+\Sigma^\Ret_{\bar t,23}$ allows to express the poles as
\begin{widetext}
\begin{align}
\omega_1 =\epsilon^{\prime\Lambda}-i(\Gamma+\Lambda) \; , \;\;\; 
\omega_{2/3} = \frac{1}{2}\left(\epsilon^\Lambda+ \epsilon^{\prime\Lambda}-i\Gamma-2i\Lambda\mp 
\sqrt{-(\Gamma-i\epsilon^\Lambda+i\epsilon^{\prime,\Lambda})^2+4|\tau_{12}^\Lambda|^2+4|\tau_{23}^\Lambda|^2}\right)~.
\end{align} 
The corresponding residues are given explicitly in Tab.\ \ref{tab:residues}.
\begin{table}
\centering
\begin{tabular}{ccccccccc}
\multicolumn{2}{c}{$\text{Res}_{ij,n}$}&\multicolumn{4}{c}{}&{\color{white}space}&$ij$&\\\noalign{\vskip0.1cm}
\multicolumn{2}{c}{}&\multicolumn{4}{c}{$ij$}&{\color{white}space}&33&{$\text{Res}_{11,n}(\tau_{12}^\Lambda\to \tau_{23}^\Lambda)$}\\\noalign{\vskip0.1cm}
\cline{8-9}
\noalign{\vskip0.1cm}
&&11&22&12&13&&21&{$\text{Res}_{12,n}(\tau_{12}^\Lambda\to (\tau_{12}^\Lambda)^*)$}\\
\noalign{\vskip0.1cm}
\cline{1-6}\cline{8-9}
\noalign{\vskip0.1cm}
&1&$1+\frac{|\tau_{12}^\Lambda|^2}{(\omega_1-\omega_2)(\omega_1-\omega_3)}$&$0$&$0$&$\frac{\tau_{12}^\Lambda \tau_{23}^\Lambda}{(\omega_1-\omega_2)(\omega_1-\omega_3)}$&&31&{$\text{Res}_{13,n}(\tau_{12}^\Lambda,\tau_{23}^\Lambda\to (\tau_{12}^\Lambda)^*,(\tau_{23}^\Lambda)^*)$}\\\noalign{\vskip0.1cm}
\cline{8-9}
\noalign{\vskip0.1cm}
n&2&$\frac{|t_{12}^\Lambda|^2}{(\omega_2-\omega_1)(\omega_2-\omega_3)}$&$\frac{\omega_2-\omega_1}{\omega_2-\omega_3}$&$\frac{\tau_{12}^\Lambda}{\omega_2-\omega_3}$&$\frac{\tau_{12}^\Lambda \tau_{23}^\Lambda}{(\omega_2-\omega_1)(\omega_2-\omega_3)}$&&23&{$\text{Res}_{12,n}(\tau_{12}^\Lambda\to \tau_{23}^\Lambda)$}\\\noalign{\vskip0.1cm}
\cline{8-9}
\noalign{\vskip0.1cm}
&3&$\frac{|\tau_{12}^\Lambda|^2}{(\omega_3-\omega_1)(\omega_3-\omega_2)}$&$\frac{\omega_3-\omega_1}{\omega_3-\omega_2}$&$\frac{\tau_{12}^\Lambda}{\omega_3-\omega_2}$&$\frac{\tau_{12}^\Lambda \tau_{23}^\Lambda}{(\omega_3-\omega_1)(\omega_3-\omega_2)}$&&32&{$\text{Res}_{23,n}(\tau_{23}^\Lambda\to (\tau_{23}^\Lambda)^*)$}\\

\end{tabular}\\ 
\caption{Residues of Eq.\ \eqref{eq:forresiduesandpoles}.}
\label{tab:residues}
\end{table}

For the integral Eq.\ (\ref{intappendix}) one substitutes the 'center of time' $T =t_1+t_2$ and the 'relative time' $\Delta t =t_2-t_1$:
\begin{equation}
\int \limits_{t_n}^{t_{n+1}}dt_1\int\limits_{m_j}^{t_{m+1}}dt_2\longrightarrow\frac{1}{2}\left[\;\int\limits_{t_m-t_{n+1}}^{t_{m+1}-t_{n+1}}d\Delta t\int\limits_{2t_m-\Delta t}^{2t_{n+1}+\Delta t}dT +\int\limits_{t_{m+1}-t_{n+1}}^{t_{m}-t_n}d\Delta t\int\limits_{2t_m-\Delta t}^{2t_{m+1}-\Delta t}dT +\int\limits_{t_m-t_{n}}^{t_{m+1}-t_n}d\Delta t\int\limits_{2t_n+\Delta t}^{2t_{m+1}-\Delta t}dT \right]
\end{equation}
for $ t_{n+1}-t_{n}\geq t_{m+1}-t_m $. The opposite case follows analogously. It proves advantageous to separate the 
problem into $  n=m $ and $  n\neq m $. We start with the case $n=m$. With the above substitution one can write the $i,j$ matrix element of Eq.\ (\ref{intappendix}) in terms of digamma $\Psi(x)$ and hypergometric functions ${}_2\mathcal{F}_1(a,b,c,z)$:\cite{abramowitz}
\begin{equation}
\begin{split}
&\left[ \int \limits_{t_n}^{t_{n+1}}ds_1\int\limits_{t_n}^{t_{n+1}}ds_2 G^{\Ret}(t_{n+1},s_1)\Sigma^\K_\res(s_1,s_2) 
G^{\Adv}(s_2,t_{n+1}) \right]_{ij}
=\lim\limits_{\delta\to 0}\sum\limits_{\alpha=L,R\atop k,l=1,2,3}T_\alpha
\text{Res}_{i \alpha,k} \text{Res}_{j \alpha,l}^* e^{-i\Delta \omega_{kl}t_{n+1}}\\
&\times\left[\;\int\limits_{t_n-t_{n+1}}^{-\delta}d\Delta t\int\limits_{2t_n-\Delta t}^{2t_{n+1}+\Delta t}dT +\int\limits_{\delta}^{t_{n+1}-t_n}d\Delta t\int\limits_{2t_n+\Delta t}^{2t_{n+1}-\Delta t}dT \right] \frac{\Gamma}{\sinh(\pi T_\alpha \Delta t) } e^{i\mu_\alpha\Delta t } e^{\frac{1}{2}i T \Delta \omega_{kl}} e^{-i\frac{1}{2}\Delta t (\omega_k+\omega_l^*)}\\
&=\sum\limits_{{\alpha=L,R} \atop{k,l}=1,2,3}\text{Res}_{i \alpha, k} \text{Res}^*_{ j \alpha,l}e^{-i(\omega_{k}t_{n+1}-\omega_l^*t_{n+1})}\frac{4T_\alpha\Gamma}{i\Delta \omega_{kl}}\bigg(e^{i\Delta\omega_{kl}t_{n+1}}\bigg[\frac{1}{2\pi T_\alpha}\left\{-\Psi\left(\frac{-i\mu_\alpha+i\omega_k+\pi T_\alpha}{2\pi T_\alpha}\right)+\Psi\left(\frac{i\mu_\alpha-i\omega_l^*+\pi T_\alpha}{2\pi T_\alpha}\right)\right\}\\
&-\frac{e^{(-i\mu_\alpha+i\omega_k+\pi T_\alpha)(t_n-t_{n+1})}}{-i\mu_\alpha+i\omega_k+\pi T_\alpha}{}_2\mathcal{F}_1\left(1,\frac{-i\mu_\alpha+i\omega_k+\pi T_\alpha}{2\pi T_\alpha},\frac{-i\mu_\alpha+i\omega_k+\pi T_\alpha}{2\pi T_\alpha}+1,e^{2\pi T_\alpha(t_n-t_{n+1})}\right)\\
&+\frac{e^{(i\mu_\alpha-i\omega_l^*+\pi T_\alpha)(t_n-t_{n+1})}}{i\mu_\alpha-i\omega_l^*+\pi T_\alpha}{}_2\mathcal{F}_1\left(1,\frac{i\mu_\alpha-i\omega_l^*+\pi T_\alpha}{2\pi T_\alpha},\frac{i\mu_\alpha-i\omega_l^*+\pi T_\alpha}{2\pi T_\alpha}+1,e^{2\pi T_\alpha(t_n-t_{n+1})}\right)\bigg]\\
&-e^{i\Delta\omega_{kl}t_{n}}\bigg[\frac{1}{2\pi T_\alpha}\left\{-\Psi\left(\frac{-i\mu_\alpha+i\omega_l^*+\pi T_\alpha}{2\pi T_\alpha}\right)+\Psi\left(\frac{i\mu_\alpha-i\omega_k+\pi T_\alpha}{2\pi T_\alpha}\right)\right\}\\
&-\frac{e^{(-i\mu_\alpha+i\omega_l^*+\pi T_\alpha)(t_n-t_{n+1})}}{-i\mu_\alpha+i\omega_l^*+\pi T_\alpha}{}_2\mathcal{F}_1\left(1,\frac{-i\mu_\alpha+i\omega_l^*+\pi T_\alpha}{2\pi T_\alpha},\frac{-i\mu_\alpha+i\omega_l^*+\pi T_\alpha}{2\pi T_\alpha}+1,e^{2\pi T_\alpha(t_n-t_{n+1})}\right)\\
\end{split}
\end{equation}
\begin{equation}
\begin{split}
&+\frac{e^{(i\mu_\alpha-i\omega_k+\pi T_\alpha)(t_n-t_{n+1})}}{i\mu_\alpha-i\omega_k+\pi T_\alpha}{}_2\mathcal{F}_1\left(1,\frac{i\mu_\alpha-i\omega_k+\pi T_\alpha}{2\pi T_\alpha},\frac{i\mu_\alpha-i\omega_k+\pi T_\alpha}{2\pi T_\alpha}+1,e^{2\pi T_\alpha(t_n-t_{n+1})}\right)\bigg]\bigg),\label{eq:GK3sites1}
\end{split}
\end{equation}
where one has exploited that for $b>0$
\begin{equation}
\lim\limits_{x\to0^+}\frac{e^{(a + b)x}}{a + b} {}_2\mathcal{F}_1\left(1,\frac{a+b}{2b},\frac{a+b}{2b}+1,e^{2bx}\right)-\frac{e^{(c + b)x}}{c + b} {}_2\mathcal{F}_1\left(1,\frac{c+b}{2b},\frac{c+b}{2b}+1,e^{2bx}\right)
  =\frac{1}{2b} \left\{-\Psi\left( \frac{ a + b}{2b}\right) + \Psi\left(\frac{c + b}{2b}\right)\right\}
   \label{eq:E1limit}~,
\end{equation}
and defined
\begin{equation}
\Delta\omega_{kl} =  \omega_k-\omega_l^*\label{eq:Deltaw}.
\end{equation}
In the indices of the residues in Eq.\ (\ref{eq:GK3sites1}) one has to replace $\alpha=L$ by $1$ and $\alpha=R$ by $3$. 
For the case $m\neq n$ one analogously finds
\begin{equation}
\begin{split}
&\left[ \int \limits_{t_n}^{t_{n+1}}ds_1\int\limits_{t_m}^{t_{m+1}}ds_2 G^{\Ret}(t_{n+1},s_1)\Sigma^\K_\res(s_1,s_2) 
G^{\Adv}(s_2,t_{m+1}) \right]_{ij}
=\frac{1}{\pi}\sum\limits_{\alpha=L,R\atop k,l=1,2,3}\text{Res}_{i\alpha,k}\text{Res}_{j\alpha,l}^* 
e^{-i(\omega_{k}t_{n+1}-\omega_l^*t_{m+1})}\\
&\times\left[\;\int\limits_{t_m-t_{n+1}}^{t_{m+1}-t_{n+1}}d\Delta t\int\limits_{2t_m-\Delta t}^{2t_{n+1}+\Delta t}dT +\int\limits_{t_{m+1}-t_{n+1}}^{t_{m}-t_n}d\Delta t\int\limits_{2t_m-\Delta t}^{2t_{m+1}-\Delta t}dT +\int\limits_{t_m-t_{n}}^{t_{m+1}-t_n}d\Delta t\int\limits_{2t_n+\Delta t}^{2t_{m+1}-\Delta t}dT \right] 
\frac{\Gamma e^{i\mu_\alpha\Delta t } e^{\frac{1}{2}i T \Delta \omega_{kl}}}{\sinh(\pi T_\alpha\Delta t) } 
e^{-i\frac{1}{2}\Delta t (\omega_k+\omega_l^*)}\\
&=\sum\limits_{{\alpha=L,R}\atop{k,l}=1,2,3}T_\alpha\text{Res}_{i\alpha,k} \text{Res}^*_{ j\alpha,l}e^{-i(\omega_{k}t_{n+1}-\omega_l^*t_{m+1})}\frac{4T_\alpha\Gamma}{i\Delta \omega_{kl}}\\
&\phantom{=}\times\bigg[-e^{i\Delta\omega_{kl}t_{n+1}}\frac{e^{(i\mu_\alpha-i\omega_l^*+\pi T_\alpha)(t_{m+1}-t_{n+1})}}{i\mu_\alpha-i\omega_l^*+\pi T_\alpha}{}_2\mathcal{F}_1\left(1,\frac{i\mu_\alpha-i\omega_l^*+\pi T_\alpha}{2\pi T_\alpha},\frac{i\mu_\alpha-i\omega_l^*+\pi T_\alpha}{2\pi T_\alpha}+1,e^{2\pi T_\alpha(t_{m+1}-t_{n+1})}\right)\\
&\phantom{=\times\bigg[}+e^{i\Delta\omega_{kl}t_{n+1}}\frac{e^{(i\mu_\alpha-i\omega_l^*+\pi T_\alpha)(t_{m}-t_{n+1})}}{i\mu_\alpha-i\omega_l^*+\pi T_\alpha}{}_2\mathcal{F}_1\left(1,\frac{i\mu_\alpha-i\omega_l^*+\pi T_\alpha}{2\pi T_\alpha},\frac{i\mu_\alpha-i\omega_l^*+\pi T_\alpha}{2\pi T_\alpha}+1,e^{2\pi T_\alpha(t_{m}-t_{n+1})}\right)\\
&\phantom{=\times\bigg[}-e^{i\Delta\omega_{kl}t_{m}}\frac{e^{(i\mu_\alpha-i\omega_k+\pi T_\alpha)(t_m-t_{n+1})}}{i\mu_\alpha-i\omega_k+\pi T_\alpha}{}_2\mathcal{F}_1\left(1,\frac{i\mu_\alpha-i\omega_k+\pi T_\alpha}{2\pi T_\alpha},\frac{i\mu_\alpha-i\omega_k+\pi T_\alpha}{2\pi T_\alpha}+1,e^{2\pi T_\alpha(t_m-t_{n+1})}\right)\\
&\phantom{=\times\bigg[}+e^{i\Delta\omega_{kl}t_{m+1}}\frac{e^{(i\mu_\alpha-i\omega_k+\pi T_\alpha)(t_{m+1}-t_{n+1})}}{i\mu_\alpha-i\omega_k+\pi T_\alpha}{}_2\mathcal{F}_1\left(1,\frac{i\mu_\alpha-i\omega_k+\pi T_\alpha}{2\pi T_\alpha},\frac{i\mu_\alpha-i\omega_k+\pi T_\alpha}{2\pi T_\alpha}+1,e^{2\pi T_\alpha(t_{m+1}-t_{n+1})}\right)\\
&\phantom{=\times\bigg[}+e^{i\Delta\omega_{kl}t_{m}}\frac{e^{(i\mu_\alpha-i\omega_k+\pi T_\alpha)(t_{m}-t_{n})}}{i\mu_\alpha-i\omega_k+\pi T_\alpha}{}_2\mathcal{F}_1\left(1,\frac{i\mu_\alpha-i\omega_k+\pi T_\alpha}{2\pi T_\alpha},\frac{i\mu_\alpha-i\omega_k+\pi T_\alpha}{2\pi T_\alpha}+1,e^{2\pi T_\alpha(t_{m}-t_{n})}\right)\\
&\phantom{=\times\bigg[}-e^{i\Delta\omega_{kl}t_{m+1}}\frac{e^{(i\mu_\alpha-i\omega_k+\pi T_\alpha)(t_{m+1}-t_{n})}}{i\mu_\alpha-i\omega_k+\pi T_\alpha}{}_2\mathcal{F}_1\left(1,\frac{i\mu_\alpha-i\omega_k+\pi T_\alpha}{2\pi T_\alpha},\frac{i\mu_\alpha-i\omega_k+\pi T_\alpha}{2\pi T_\alpha}+1,e^{2\pi T_\alpha(t_{m+1}-t_{n})}\right)\\
&\phantom{=\times\bigg[}-e^{i\Delta\omega_{kl}t_{n}}\frac{e^{(i\mu_\alpha-i\omega_l^*+\pi T_\alpha)(t_{m+1}-t_{n})}}{i\mu_\alpha-i\omega_l^*+\pi T_\alpha}{}_2\mathcal{F}_1\left(1,\frac{i\mu_\alpha-i\omega_l^*+\pi T_\alpha}{2\pi T_\alpha},\frac{i\mu_\alpha-i\omega_l^*+\pi T_\alpha}{2\pi T_\alpha}+1,e^{2\pi T_\alpha(t_{m+1}-t_{n})}\right)\\
&\phantom{=\times\bigg[}-e^{i\Delta\omega_{kl}t_{n}}\frac{e^{(i\mu_\alpha-i\omega_l^*+\pi T_\alpha)(t_{m}-t_{n})}}{i\mu_\alpha-i\omega_l^*+\pi T_\alpha}{}_2\mathcal{F}_1\left(1,\frac{i\mu_\alpha-i\omega_l^*+\pi T_\alpha}{2\pi T_\alpha},\frac{i\mu_\alpha-i\omega_l^*+\pi T_\alpha}{2\pi T_\alpha}+1,e^{2\pi T_\alpha(t_{m}-t_{n})}\right)
\bigg].\label{eq:GK3sites1_a}
\end{split}
\end{equation}
Similarly one can derive $S^\K(t,t)$, which additionally involves the trigamma function $\Psi(1,x)$. 
\end{widetext}




\begin{thebibliography}{99}
\bibitem{comment1} Without any approximations the high energy cutoff $\Gamma$ appears naturally. In the derivation of the analytical Eq.~\ref{eq:cutoffbeforeint} we neglected terms of the order $1/\Gamma^2$ which now makes it necessary to introduce the high energy cutoff as an upper bound of integration by hand.
\bibitem{comment2} Using the noninteracting version of the Meir-Wingreen expression for the current does not constitute an additional approximation within our approach since at the end of the RG flow we have to compute the current of a renornmalized noninteracting system.
\end{thebibliography}
\end{document}